\documentstyle{article} 
\title{Algorithmic Chaos}

\author{
Paul M.B. Vit\'{a}nyi\thanks{This paper is based
on a talk by the author at the University of Waterloo, Canada, in 1991.
Partially supported by EU through NeuroColt II Working Group 
and the QAIP Project.
Address: Centrum voor Wiskunde en Informatica,
Kruislaan 413, 1098 SJ Amsterdam, The Netherlands. Email: paulv@cwi.nl}\\
CWI and Universiteit van Amsterdam}

\newtheorem{theo}{Theorem}
\newtheorem{lemma}{Lemma}

\newtheorem{definition}{Definition}

\newenvironment{proof}{\par \sc Proof. \rm}{$\Box$ \vspace{1ex}}

\newcommand{\ar}{\rightarrow}

\newcommand{\half}{\frac{1}{2}}

\newcommand{\defic}{\delta}
\newcommand{\C}{C}
\newcommand{\K}{K}

\date{}
\begin{document}

\maketitle

\begin{abstract}
Many physical theories like
chaos theory are fundamentally concerned with the
conceptual tension between determinism and randomness. 
Kolmogorov complexity can express randomness in determinism
and gives an approach to formulate
chaotic behavior.
\end{abstract}

\section{Introduction}

Ideally, physical theories are abstract representations---mathematical
axiomatic theories for the underlying
physical reality. This reality cannot be directly
experienced, and is therefore unknown and even
in principle unknowable. Instead, scientists postulate
an informal description which is intuitively acceptable,
and subsequently formulate one or more mathematical
theories to describe the phenomena.

{\bf Deterministic Chaos:} Many phenomena in physics (like the weather) satisfy
well accepted deterministic equations. From initial
data we can extrapolate and compute the next states
of the system. Traditionally it was thought that increased
precision of the initial data (measurement) and increased
computing power would result in increasingly accurate
extrapolation (prediction). Unfortunately it turns out
that for many systems this is not the case. In fact,
it turns out that any long range prediction with
any confidence better than what we would get by flipping a fair
coin is practically impossible: this phenomenon is known as chaos
(see \cite{De89} for an introduction). There are two, more or less
related, causes for this:

\begin{description}
\item[Instability]
In certain deterministic systems,
an arbitrary small error in initial conditions can exponentially
increase during the subsequent evolution of the system,
until it encompasses the full range of values achievable
by the system.
This phenomenon of instability of a computation is in fact
well known in numerical analysis: computational procedures
inverting ill-conditioned matrices (with determinant about zero)
will introduce exponentially increasing errors. 
\item[Unpredictability]
Assume we deal with a system described by deterministic equations
which can be finitely represented (like a recursive function).
Even if fixed-length initial segments of the 
infinite binary representation of the real parameters
describing  past states of the system are perfectly known,
and the computational procedure used is perfectly error free,
for many such systems it will still be 
impossible to effectively predict (compute) any
significantly long extrapolation of system states with any confidence
higher than using a random coin flip. This is the 
core of chaotic phenomena: randomness in determinism.
\end{description}

{\bf Probability:}
Classical probability theory deals with randomness in the
sense of {\em random variables}. The concept of random individual
data cannot be expressed. Yet our intuition about the latter is
very strong:
An adversary claims to have a true random coin and invites
us to bet on the outcome. The coin produces a hundred heads
in a row. We say that the coin cannot have been fair. The adversary,
however, appeals to probability theory which says that each sequence
of outcomes of a hundred coin flips is equally likely, $1/2^{100}$,
and one sequence had to come up. 
Probability theory gives us no basis to challenge an outcome
{\em after} it has happened. We could only exclude unfairness
in advance by putting a penalty side-bet on an outcome
of 100 heads. But what about $1010 \ldots$? What about
an initial segment of the binary expansion of $\pi$?

\begin{description}
\item[Regular sequence]
\[ \Pr(00000000000000000000000000) = \frac{1}{2^{26}} \]
\item[Regular sequence]
\[ \Pr(01000110110000010100111001) = \frac{1}{2^{26}} \]
\item[Random sequence]
\[ \Pr(10010011011000111011010000) = \frac{1}{2^{26}} \]
\end{description}

The first sequence is regular, but what is the distinction of
the second sequence and the third? The third sequence was generated
by flipping a quarter. The second sequence is very regular:
$0,1,00,01, \ldots $. The third sequence will pass
(pseudo) randomness tests.

In fact, classical probability theory cannot express
the notion of {\em randomness of an individual sequence}.
It can only express expectation of properties of the
total set of sequences under some distribution.

This is analogous to the situation in physics
above. How can `an individual object be random?'
is as much a probability theory paradox as
`how can an individual sequence
of states of a deterministic system be random?'
is a paradox of deterministic physical systems.

In probability theory the problem has found a satisfactory
resolution by combining notions of computability 
and information theory to express the complexity of a finite object.
This complexity is the length of the shortest
binary program from which the object can be effectively
reconstructed. It may be called the {\em algorithmic
information content} of the object. This quantity turns
out to be an attribute of the object alone, and recursively
invariant. It is the {\em Kolmogorov complexity} of the
object. It turns out that this notion can be brought to bear on the physical
riddles too.

\section{Kolmogorov Complexity}

\label{sec.complexity}
To make this paper
self-contained we briefly review notions and properties required.
For details and further properties see the textbook \cite{LV97}.
We identify the natural numbers ${\cal N}$ and the finite binary sequences as
\[( 0, \epsilon ), (1,0), (2,1),(3,00),(4,01), \ldots ,\]
where $\epsilon$ is the empty sequence.
The {\em length} $l(x)$ 
of a natural number $x$ is the number of
bits in the corresponding binary sequence.
For instance, $l( \epsilon ) = 0$.
If $A$ is a set, then $|A|$ denotes the {\em cardinality} of $A$.
Let $\langle . \rangle : {\cal N} \times {\cal N} \ar {\cal N}$ denote a standard
computable bijective `pairing' function. Throughout this paper, we will assume
that $\langle x,y \rangle  = 1^{l(x)}0xy$.

Define $\langle x,y,z \rangle $ by $\langle x, \langle y,z \rangle  \rangle $. 

We need some notions from the theory of algorithms, see \cite{Ro67}.
Let $\phi_1 , \phi_2 , \ldots $ be a standard enumeration of the
partial recursive functions.
The (Kolmogorov) {\em  complexity} of $x \in {\cal N}$, given $y$, is defined as

\[ \C(x| y) = \min \{ l(\langle n,z \rangle ): \phi_n (\langle y,z \rangle ) = x \} .\]

This means that $\C(x|y)$ is the {\em minimal} number of bits 
in a description
from which $x$ can be effectively reconstructed, given $y$.
The unconditional complexity is defined as $\C(x)=\C(x| \epsilon )$.

An alternative definition is as follows. Let
\begin{equation}
\label{universal}
 \C_{\psi}(x|y) = \min \{ l(z): \psi (\langle y,z \rangle ) = x \}
\end{equation} 
be the conditional
complexity of $x$ given $y$ with reference to decoding function $\psi$.
Then $\C(x|y) = \C_{\psi}(x|y)$
for a universal partial recursive function $\psi$
that satisfies $\psi (\langle y,n,z \rangle )= \phi_{n} (\langle y,z \rangle )$.

We will also make use of the {\em prefix\/} complexity $\K(x)$, which denotes
the shortest {\em self-delimiting\/} description.
To this end, we consider so called {\em prefix\/} Turing machines,
which have only 0's and 1's on their input tape,
and thus cannot detect the end of the input.
Instead we define an input as that part of the input tape which the machine
has read when it halts. When $x \neq y$ are two such input, we clearly
have that $x$ cannot be a prefix of $y$, and hence the set of inputs forms
what is called a {\em prefix code\/}.
We define $\K(x)$ similarly as above, with reference to a universal
prefix machine that first reads $1^n0$ from the input tape and then
simulates prefix machine $n$ on the rest of the input.

We need the following properties. 
Throughout `$\log$' denotes the binary logarithm.
We often use $O(f(n)) = - O(f(n))$, so that 
$O(f(n))$ may denote a negative quantity.
For each $x,y\in {\cal N}$ we have 
\begin{equation}
\C(x|y) \leq l(x) + O(1).
\label{upperbound}
\end{equation}
For each $y \in {\cal N}$ there is an $x \in {\cal N}$ of length $n$
such that $\C(x|y) \geq n$. In particular, we can set $y = \epsilon$.
Such $x$'s may be called {\em random}, since they are
without regularities that can be used to compress
the description. Intuitively,
the shortest effective description of $x$
is $x$ itself. In general, for each $n$ and $y$, there are at least
$2^n -2^{n-c} +1$ distinct $x$'s of length $n$ with
\begin{equation}
\C(x|y) \geq n-c.
\label{random}
\end{equation}

In some cases we want to encode $x$ in {\em self-delimiting}
form $x'$, in order to be able to decompose $x'y$
into $x$ and $y$. 
Good upper bounds on the prefix complexity of $x$ are obtained by iterating
the simple rule that a self-delimiting (s.d.) description of the length of $x$
followed by $x$ itself is a s.d. description of $x$. 
For example, $x'=1^{l(x)} 0 x$ and $x''=1^{l(l(x))}0l(x)x$ are
both s.d. descriptions for $x$, and this shows
that $\K(x) \leq 2l(x)+O(1)$ and $\K(x) \leq l(x) + 2l(l(x))+ O(1)$.

Similarly, we can encode $x$
in a self-delimiting form of its shortest program $p(x)$
($l(p (x))=\C(x)$) in $2\C(x) + 1$ bits.
Iterating this scheme, we can encode $x$ as
a selfdelimiting program of $\C(x)+2 \log \C(x) +1$ bits, 
which shows that $\K(x) \leq \C(x)+2 \log \C(x) +1$,
and so on.

The string $sqi$ has length at most $n - \defic(n) - O(1)$ and can be padded

\subsection{Random Sequences}

We would like to call an infinite sequence $\omega \in \{0,1\}^{\infty}$
random if $\C ( \omega_{1:n} ) \geq n - O(1)$ for all $n$.
It turns out that such sequences do not exist. This occasioned
the celebrated theory of randomness of P. Martin-L\"of,
\cite{Ma66}. Later it turned out, \cite{Ch75}, that we can yet precisely
define the Martin-L\"of random sequences, but using
prefix Kolmogorov complexity.
we need.

\begin{theo}
\label{martin.lof}
An infinite binary sequence $\omega$ is random in the sense of Martin-L\"of 
iff there is an $n_0$ such that
$\K( \omega_{1:n} ) \geq n$ for all $n > n_0$,
\end{theo}

That $\omega$ is random in Martin-L\"of's sense means that
it will pass {\em all} effective tests for randomness:
both the tests which are known now and the ones which are
as yet unknown \cite{Ma66}. 

Similar properties hold for high-complexity finite strings,
although in a less absolute sense.

For every finite set $S  \subseteq \{0,1\}^*$ containing
$x$ we have $K(x | S)\le\log|S|+O(1)$.
Indeed, consider the selfdelimiting code of $x$
consisting of its $\lceil\log|S|\rceil$ bit long index
of $x$ in the lexicographical ordering of $S$.
This code is called
\emph{data-to-model code}.
The lack of typicality
of $x$ with respect to $S$
is the amount by which $K(x|S)$
falls short of the length of the data-to-model code.
The {\em randomness deficiency} of $x$ in $S$ is defined by
      \begin{equation}\label{eq:randomness-deficiency}
\delta (x | S) = \log |S| - K(x | S),
      \end{equation}
for $x \in S$, and $\infty$ otherwise.
If $\delta(x | S)$ is small, then $x$ may be considered
as a {\em typical} member
of $S$. 
There are no simple special properties that
single it out from the majority of elements in $S$.
This is not just terminology: If $\delta (x | S)$ is small, then $x$
satisfies {\em all} properties of low Kolmogorov complexity
that hold with high probability for the elements of $S$. For example:
Consider strings $x$ of length $n$ and let $S = \{0,1\}^n$ be a set of such
strings. Then $\delta (x | S) = n - K(x |ni \pm O(1)$.

(i) If $P$ is a property satisfied by all $x$ with
$\delta(x | S) \le \delta (n)$,
then $P$ holds with probability at
least $1-1/2^{\delta(n)}$ for the elements of $S$.

(ii) Let
$P$ be any
property
that holds with probability at least
$1-1/2^{\delta (n)}$ for the
elements of $S$. Then, every such $P$ holds
simultaneously for every $x \in S$
with $\delta (x | S)\le\delta (n)-K(P|n)-O(1)$.

\section{Algorithmic Chaos Theory}

For convenience assume that time is discrete: ${\cal N}$.
In a deterministic system $X$ the 
{\em state} of the system at time $t$ is $X_t$.
The {\em orbit} of the system is the sequence of subsequent states
 $X_0 , X_1 , X_2 , \ldots$. For convenience we assume
the states are elements of $\{0,1\}$. The definitions below
are easily generalized. For each system, be it deterministic
or random, we associate a measure $\mu$ with the space
$\{0,1\}^{\infty}$ of orbits. That is, $\mu (x)$
is the probability that an orbit starts with $x \in \{0,1 \}^*$.

Given an initial segment $X_{0:t}$ of the orbit we want
to compute $X_{t+1}$. Even if it would not be possible
to compute $X_{t+1}$, we would like to compute a 
prediction of it which does better than a random coin flip.

\begin{definition}\label{def.chaos}
\rm
Let the set of orbits be $S= \{0,1\}^{\infty}$
with the Lebesgue measure $\lambda$.
Let $\phi$ be a partial recursive function and 
let $\omega \in S$. Define
\[ \zeta_i = \left\{ \begin{array}{ll}
1 & \mbox{if $\phi ( \omega_{1:i-1} ) = \omega_i$} \\
0 & \mbox{otherwise}
\end{array}
\right. \]
A deterministic system is {\em chaotic} if, for every
computable function $\phi$, we have
\[ \lim_{t \rightarrow \infty} \sum_{i=0}^{t-1} \zeta_i 
 = \frac{1}{2} ,\]
with probability 1.
\end{definition}

A well-known example of a chaotic system is the {\em doubling map},
\cite{Fo83}. 
Consider the {\it deterministic} system $X$ with initial state
$X_0 = 0. \omega$ a real number in the interval $[0,1]$
where $\omega \in S$ is the binary representation. 
\begin{equation}\label{ford.1}
X_{n+1} = 2X_n \pmod{1} 
\end{equation}
where $\pmod{1}$ means drop the integer part. Thus, all iterates of
equation~\ref{ford.1} lie in the unit interval $[0,1]$. In physics,
this is called the `energy surface' of the orbit.
We can partition this energy surface into
two cells, a left cell $[0, \frac{1}{2} )$ and a right cell
$[\frac{1}{2} , 1]$. Thus $X_n$ lies in the left cell
if and only if the $n$th digit of $\omega$ is 0.

One way to derive the doubling map is as follows:
In chaos theory, \cite{De89}, people have for
years being studying the discrete logistic equation
\[ Y_{n+1} = \alpha Y_n (1 - Y_n ) \]
which maps the unit interval upon itself when $0 \leq \alpha \leq 4$.
When $\alpha =4$, setting $Y_n = \sin^2 \pi X_n$,
we obtain:
\[ X_{n+1} = 2 X_n \pmod{1} . \]

\begin{lemma}
There are a chaotic systems (like $X$ and $Y$ above).
\end{lemma}
\begin{proof}
We prove that $X$ is a chaotic system. Since $Y$ reduces to $X$
by specialization,this shows that $Y$ is chaotic as well.
Assume $\omega$ is random. Then by Theorem~\ref{martin.lof}, 
\begin{equation}\label{eq.omega.random}
C( \omega_{1:n} ) > n- 2 \log n +O(1).
\end{equation}
Let $\phi$ be any partial recursive function.
Construct $\zeta$ from $\phi$ and $\omega$ as in Definition~\ref{def.chaos.finite}.

Assume by way of contradiction that there is an $\epsilon > 0$ such that
\[ | \frac{1}{n} \lim_{n \rightarrow \infty} 
\sum_{i=1}^n \zeta_i - \frac{1}{2} | \geq \epsilon . \]
Then,
there is a $\delta > 0$ such that
\begin{equation}\label{eq.zeta.compl}
\lim_{n \rightarrow \infty} \frac{C( \zeta_{1:n} )}{n} \leq (1- \delta ) . 
\end{equation}

We prove this as follows. The number of binary sequences of length $n$
where the numbers of 0's and 1's differ by at least an $\epsilon n$ 
is 
\[ N = 2 \cdot 2^n \sum_{m = ( \half + \epsilon)n}^n b(n, m, \half )\]
where $b(n,m,p)$
is the probability of $m$ successes out of $n$ trials in a
$( p , 1-p )$ Bernoulli process: the Binomial distribution.
A general estimate of the tail probability of the binomial distribution,
with $m$ the number of successful outcomes in $n$ experiments
with probability of success $0 < p < 1$ and $q=1-p$, 
is given by Chernoff's
bounds, \cite{ES74,CLR90},
\begin{equation}
\Pr (|m -np| \geq \epsilon n ) \leq 2e^{- ( \epsilon n)^2 /3n } .
\label{chernoff}
\end{equation}
Therefore, we can describe any element $\zeta_{1:n}$
concerned by giving
$n$ and $\epsilon n$ in $2 \log n + 4 \log \log n$ bits selfdelimiting
descriptions, and pointing out the string concerned
in a constrained ensemble of at most $N$ elements in $\log N$ bits.
Therefore,
\[ C(\zeta_{1:n}) \leq n - \epsilon^2 n \log e + 2 \log n +
4 \log \log n + O(1) . \]
That is, we can choose
\[ \delta = \epsilon^2 \log e + 
\frac{2 \log n + 4 \log \log n + O(1)}{n} . \]
Next, given $\zeta$ and $\phi$ we can reconstruct $\omega$
as follows:

\tt 
for $i := 1,2, \ldots$ do: \\
	if $\phi ( \omega_{1:i-1} ) = a$ and $\zeta_i = 0$ then
	$\omega_i := \neg a$ \\
	else $\omega_i := a$.
\rm

Therefore,
\begin{equation}\label{eq.omega.compl}
C( \omega_{1:n} ) \leq C( \zeta_{1:n} ) + K ( \phi ) + O(1) .
\end{equation}
Now Equations~\ref{eq.omega.random}, \ref{eq.zeta.compl},
\ref{eq.omega.compl} give the desired contradiction.
By Theorem~\ref{martin.lof}, the set of $\omega$'s satisfying
Equation~\ref{eq.omega.random} has uniform measure one,
which proves the lemma.
\end{proof}

 In \cite{Fo83} the argument is as follows.
 Assuming that the initial state is randomly drawn from $[0,1)$
 according to the uniform measure $\lambda$, we can use 
complexity arguments to show that the doubling map's observable
 orbit cannot be predicted better than a coin toss.
 Namely, with $\lambda$-probability 1 the drawn
 initial state will be a Martin-L\"of random
 infinite sequence. Such sequences
 by definition cannot be effectively predicted
 better than
 a random coin toss, see \cite{Ma66}.

 But in this case we do not need to go
 to such trouble. The observed orbit essentially
 consists of the consecutive bits of the initial state.
 uniform measure is isomorphic to flipping a fair
 coin to generate it. This raises the challenging problem
 of a meaningful application of Kolmogorov complexity
 to chaos problems.

From a practical viewpoint it may be argued that we really
are not interested in infinite sequences: in practice
the input will always be finite precision. Now an infinite
sequence which is random may still have an arbitrary long finite
initial segment which is completely regular. Therefore,
we analyse the theory for finite precision inputs in the
following section. 

\subsection{Chaos with Finite Precision Input}

In the case of infinite precision real inputs, the distinction
between chaotic and non-chaotic systems can be precisely
drawn. In the case of finite precision inputs the distinction
is necessarily a matter of degree. This occasions
the following definition.

\begin{definition}\label{def.chaos.finite}
\rm
Let $S, \lambda , \phi , \omega$ and 
$\zeta$ be as in Definition~\ref{def.chaos}.
A deterministic system with {\em input
precision} $n$ is $( \epsilon , \delta )$-{\em chaotic} if, 
for every
computable function $\phi$, we have
\[ |\sum_{i=1}^{n} \zeta_i 
 - \frac{1}{2} | \geq \epsilon ,\]
with probability at least $1- \delta$.
\end{definition}

So systems are chaotic in the sense
of Definition~\ref{def.chaos}, like the doubling map above,
iff they are $(0,0)$-chaotic with precision $\infty$.
The system is {\em probably approximately inpredictable}:
a {\em pai}-chaotic system.

%

\begin{theo}
Systems $X$ and $Y$ above are $( \sqrt{ (\delta (n) + O(1)) \ln 2 /n},
1/2^{\delta (n)}$-chaotic for every function $\delta$ such that
$0 < \delta(n) < n$ .

\end{theo}

\begin{proof}
We prove that $X$ is $(\epsilon, \delta)$-chaotic. Since $Y$ reduces to $X$,
this implies that $Y$ is $(\epsilon, \delta)$-chaotic as well.

Assume that $x$ is a binary string of length $n$ with
\begin{equation}\label{eq.random.x}
C(x) \geq n- \delta (n) .
\end{equation}
Let $\phi$ be a polynomial time computable function, and define $z$ by:
\[ z_i = \left\{ \begin{array}{ll}
1 & \mbox{if $\phi ( x_{1:i-1} ) = x_i$} \\
0 & \mbox{otherwise}
\end{array}
\right. \]

Then, $x$ can be reconstructed from $z$ and $\phi$ as before,
and therefore:
\[ C(x) \leq C(z) + K( \phi ) + O(1) . \]
By Equation~\ref{eq.random.x} this means

\begin{equation}\label{eq.char.z}
C(z) \geq n - \delta (n) - K( \phi ) + O(1) . 
\end{equation}

We analyse the number of zeros and ones in $z$.
Using Chernoff's bounds,  Equation~\ref{chernoff},
with 
$p=q= \frac{1}{2}$,
the number of $z$'s which have an excess of $\epsilon n$
of ones over zeros is:

\[ N \leq 2^{n+1}e^{- ( \epsilon n )^2 / n } \]

with
\[ |\#ones(x)- \frac{n}{2} | < \epsilon n .\]

Then, we can give an effective description of $z$ by giving a description of
$\phi$, $\delta$ and $z$'s index in the set of size $N$
in this many bits
\begin{equation}
\label{eq.descr.z}
 n - \epsilon^2 n \log e + K( \phi ) + K ( \delta ) + 
2 \log  K( \phi ) K ( \delta ) + O(1).
\end{equation}

From Equations~\ref{eq.char.z}, \ref{eq.descr.z} we find 
\begin{equation}
\epsilon \leq \frac{ \sqrt{ \delta (n) + 2K( \phi )
+ K( \delta ) + 2 \log K( \phi ) K ( \delta ) + O(1)}}{n \log e}
\end{equation}

Making the simplifying assumption that $K( \phi ), K( \delta ) = O(1)$
this yields
\begin{equation}\label{result1}
 |\#ones(z)- \frac{n}{2} | < \sqrt{ ( \delta (n) + O(1))n \ln 2 }
\end{equation}

The number of binary strings $x$ of length $n$ with
$C(x) < n - \delta (n)$ is at most $2^{n- \delta (n)} -1$
(there are not more programs of length less than $n- \delta (n)$).
Therefore, the uniform probability of a real number starting
with an $n$-length initial segment $x$ such that $C(x) \geq n - \delta (n)$
is given by: 

\begin{equation}\label{eq.prob}
\lambda \{ \omega : C( \omega_{1:n} \geq n - \delta (n) \}
> 1 - \frac{1}{2^{\delta (n)}} . 
\end{equation}

Therefore, system $X$ is $( \epsilon , \delta )$ chaotic
with $\epsilon = \sqrt{ (\delta (n) + O(1)) \ln 2 /n}$
and $\delta = 1/2^{\delta (n)}$.

\end{proof}

\end{document}